**Thermally induced stresses in boulders on airless body surfaces, and implications for rock breakdown**

J.L. Molaro[1*], S. Byrne[2], J.-L. Le[3]

[1]Jet Propulsion Laboratory, California Institute of Technology, 4800 Oak Grove Drive, Pasadena, CA 91109, USA.

[2]Lunar and Planetary Laboratory, University of Arizona, 1629 E University Blvd, Tucson, AZ 85721, USA

[3]Department of Civil, Environmental, and Geo Engineering, University of Minnesota, 500 Pillsbury Drive, Minneapolis, MN 55455, USA

* Corresponding author: Phone +1 818 393 3087

email: jmolaro@jpl.nasa.gov

Abstract

This work investigates the macroscopic thermomechanical behavior of lunar boulders by modeling their response to diurnal thermal forcing. Our results reveal a bimodal, spatiotemporally-complex stress response. During sunrise, stresses occur in the boulders' interiors that are associated with large-scale temperature gradients developed due to overnight cooling. During sunset, stresses occur at the boulders' exteriors due to the cooling and contraction of the surface. Both kinds of stresses are on the order of 10 MPa in 1 m boulders and decrease for smaller diameters, suggesting that larger boulders break down more quickly. Boulders ≤30 cm exhibit a weak response to thermal forcing, suggesting a threshold below which crack propagation may not occur. Boulders of any size buried by regolith are shielded from thermal breakdown. As boulders increase in size (>1 m), stresses increase to several 10s of MPa as the behavior of their surfaces approaches that of an infinite halfspace. As the thermal wave loses contact with the boulder interior, stresses become limited to the near-surface. This suggests that the survival time of a boulder is not only controlled by the amplitude of induced stress, but also by its diameter as compared to the diurnal skin depth. While stresses on the order of 10 MPa are enough to drive crack propagation in terrestrial environments, crack propagation rates in vacuum are not well constrained. We explore the relationship between boulder size, stress, and the direction of crack propagation, and discuss the implications for the relative breakdown rates and estimated lifetimes of boulders on airless body surfaces.

1. Introduction

Thermally induced rock breakdown is thought to be an active process in the solar system, contributing to crater degradation and regolith production on planetary surfaces. This process is driven by the propagation of microcracks due to complex spatiotemporal stress fields induced in boulders at both micro- and macroscopic scales (Kranz, 1983; Molaro et al., 2015). The slow accumulation and propagation of microcracks in boulders over time develops macroscopic-scale features, and contributes to disaggregation of rocky material and an overall reduction in strength and elastic modulus. While this has been demonstrated by terrestrial field studies (Eppes et al., 2016; Warren et al., 2013), and laboratory studies of thermal cycling in Earth (Jansen et al., 1993; Kranz, 1983; Simmons and Cooper, 1978) and Mars (Viles et al., 2010) atmospheric environments, this topic has also become of interest in the context of airless body surfaces. Strong thermally induced stresses occur on bodies with large diurnal temperature ranges, and thus those with small solar distances and/or long day lengths (Molaro and Byrne, 2012; Molaro et al., 2015), suggesting that many airless bodies may be particularly susceptible to this "fatigue" process. Laboratory studies have shown that thermal cycling can fracture rocks in lunar environment (Thirumalai and Demou, 1970), and ordinary, CM, and LL/L chondrites in atmosphere (Delbo et al., 2014; Levi, 1973), however research in this area is still limited for planetary applications. Thermal breakdown has been suggested to play an active role in the surface evolution of the Moon (Molaro et al., 2015), Mercury (Molaro and Byrne, 2012), near-Earth asteroids (Delbo et al., 2014; Dombard et al., 2010; Jewitt and Li, 2010), and even comets(Alí-Lagoa et al., 2015; Gulkis et al., 2015; Maarry et al., 2015), emphasizing the broad significance it could have in our understanding of the evolution of airless body surfaces.

There are many active processes on airless bodies that contribute to rock breakdown and regolith production, such as impact cratering, space weathering, and micrometeorite bombardment. Micrometeorite bombardment has been assumed to dominate regolith generation on the Moon, with an estimated survival time for centimeter- to meter-sized rocks of $10^6$-$10^8$ years (Basilevsky et al., 2013; Ghent et al., 2014; Hörz and Cintala, 1997). However, breakdown rates due to thermal stresses are not well constrained, and how they compare to this estimate is unknown. Quantifying their contribution to rock breakdown on the Moon could have important implications for estimates of surface and crater ages. On Mercury, the maturation rate of regolith is four times faster than on the Moon (Braden and Robinson, 2013), in part because of its higher macro- and micro-scale impactor flux (Cintala, 1992; Le Feuvre and Wieczorek, 2011). However, Mercury's extreme diurnal temperature range suggests that thermal breakdown could make a significant contribution to this process. Asteroids typically have very small diurnal temperature ranges compared to the Moon and Mercury, however many have very rapid rotation rates. If thermal breakdown is active on these surfaces, it could be a very efficient process even if only a small amount of crack propagation occurs during each cycle.

Even if a single weathering process dominates regolith production on a given body, all processes will interact with each other to some degree. Thus better understanding how thermal breakdown operates on these surfaces will inform our understanding of regolith production rates in broader terms. For example, Ghent et al.

(2014) and Bandfield et al. (2011) found that while most of the Moon's surface has a negligible boulder population, ejecta blankets surrounding young impact craters have significant numbers of meter sized rocks that have not yet broken down or become buried. Additionally, mass wasting from steep crater walls can replenish rocks at the surface (Ghent et al., 2014; Xiao et al., 2013). Both of these processes provide source material for thermal breakdown, and may affect not only regolith production rates, but also its spatial distribution. Similarly, micrometeorite bombardment creates fresh surfaces to weather at microscopic scales, and may emplace impact damage for thermal stresses to exploit.

In order to address these broader questions, the work presented here, and that of Molaro et al. (2015), attempts to develop a more fundamental understanding of how thermal breakdown operates as a process. Molaro et al. (2015), modeled diurnal temperatures and stresses of the lunar surface at the microstructure scale, and found that the magnitude of induced stress is controlled by grain heterogeneity, with strong stresses developing along surface-parallel grain boundaries between minerals with differing elastic properties. The nature of breakdown at this scale is controlled by the grain distribution, as the complex stress field set up across and between multiple grains determines the path a crack takes as it propagates. Their work specifically isolated grain-scale processes, and did not address the macroscopic effects of the size and shape on an object, such as a lunar boulder. The boulder's size and shape will determine the amount of radiation received from the sun, radiation exchange with surrounding regolith and topography, and the efficiency of thermal emission from its surface. Throughout the course of the day, the passing thermal wave will develop a complex spatiotemporal temperature and stress field within the boulder, that will interact with interact with grain scale processes to drive crack growth. Thus, in this study we model the thermomechanical response of boulders of varying size on the surface of the Moon and other airless bodies to diurnal thermal forcing. We will explore the magnitude and macro-scale distribution of induced stresses, and discuss the implications for breakdown rates and lifetimes of boulders on these surfaces.

2. Model

In this study, we model temperatures and stresses in boulders on airless body surfaces using COMSOL Multiphysics, a commercially available 3D finite element simulation program. The geometry of our model consists of a sphere of diameter $D$ embedded in a rectangular volume of regolith (Figure 1). The regolith volume has a length and width of 4D, which was found sufficiently incorporate the effects of radiation exchange between the rock and regolith (see Appendix A.1). For $D \leq 1\ m$, the regolith volume has a depth of 4.5 m, which is 5 times the diurnal skin depth of rock on the Moon (~0.8 m). Larger boulders are modeled with deeper regolith to accommodate their size. A special case is also presented, in which a 0.1 m sphere is entirely buried beneath the regolith surface. The resolution of the finite element mesh in each case chosen to balance accuracy with computational efficiency, and is reflected in the uncertainty values presented in the results (see Appendix A.1).

The material properties for the rocks and regolith are provided in Tables 1 and 2. The rock is assigned properties of basalt, a common rock type found on the moon, and sensitivity to these properties is explored in section 3.4. Most of these material properties are constant, however heat capacity has strong temperature dependence. We follow the example of Ledlow et al. (1992), defining the heat capacity ($c_p$) as a piece-wise function of temperature (T) in units of cal/(g K) given by:

$$c_p(T \leq 350\ K) = 0.1812 + 0.1191\left(\frac{T-300}{300}\right) + 0.0176\left(\frac{T-300}{300}\right)^2 + 0.2721\left(\frac{T-300}{300}\right)^3 + 0.1869\left(\frac{T-300}{300}\right)^4$$

$$c_p(T > 350\ K) = 0.2029 + 0.0383\left(1 - \exp\left(\frac{300-T}{300}\right)\right) \quad (1)$$

This function has been shown to be accurate for lunar materials over a large temperature range and has been widely used in previous studies (Vasavada et al., 2012). We assign both the rock and the regolith in our model the heat capacity given by (1), which has an average value over the lunar temperature range of ~730 (J/kg K).

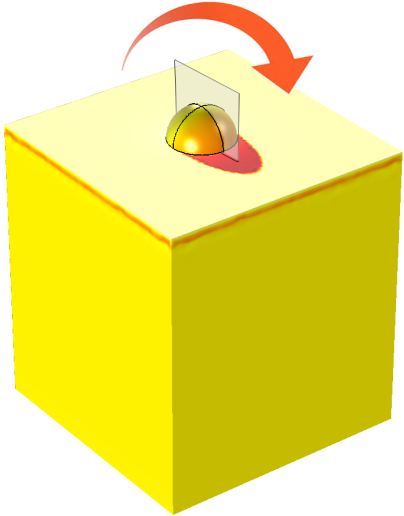

Figure 1. Geometry of a 1 m boulder embedded in a regolith block 8 m in width and height, and 4.5 m in depth. The color in the image shows the temperature of the rock and regolith in the afternoon. The arrow indicates the motion of the sun, and the semi-opaque rectangle indicates the location and orientation of the 2D plane shown in Figures 3, 4, and 8.

Table 1: Material properties for basalt with corresponding references.

| Basalt | Symbol | Units | Value | Reference |
|---|---|---|---|---|
| Thermal Conductivity | $k_B$ | W/m K | 2 | (Turcotte and Schubert, 2014) |
| Density | $\rho_B$ | kg/m$^3$ | 3150 | (Kiefer et al., 2012) |
| Surface Emissivity | $\varepsilon_B$ |  | 0.9 | (Bandfield et al., 2011) |
| Heat Capacity | $c_p$ | J/kg K | Eq. 1 | (Ledlow et al., 1992) |
| Young's Modulus | $E_B$ | GPa | 50 | (Turcotte and Schubert, 2014) |
| Poisson's ratio | $\nu_B$ |  | 0.23 | (Turcotte and Schubert, 2014) |
| Coefficient of Expansion | $\alpha_B$ | 1/K | $1*10^{-5}$ | (Turcotte and Schubert, 2014) |

Table 2: Material properties for regolith with corresponding references.

| Regolith | Symbol | Units | Value | Reference |
|---|---|---|---|---|
| Thermal Conductivity | $k_R$ | W/m K | Eq. z3 | (Vasavada et al., 2012) |
| Surface Conductivity | $k_s$ | W/m$^2$ | 0.007 | (Vasavada et al., 2012) |
| Depth Conductivity | $k_d$ | W/m$^2$ | 0.0006 | (Vasavada et al., 2012) |
| Radiative Conduction Ratio | $\chi$ | | 2.7 | (Vasavada et al., 2012) |
| Density | $\rho_R$ | kg/m$^3$ | Eq. z2 | (Vasavada et al., 2012) |
| Surface Density | $\rho_s$ | kg/m$^3$ | 1300 | (Vasavada et al., 2012) |
| Depth Density | $\rho_d$ | kg/m$^3$ | 1800 | (Vasavada et al., 2012) |
| Surface Emissivity | $\varepsilon_R$ | | 0.95 | (Bandfield et al., 2011) |
| Heat Capacity | $c_p$ | J/kg K | Eq. z1 | (Ledlow et al., 1992) |
| Young's Modulus | $E_R$ | MPa | 8 | (Colwell et al., 2007) |
| Poisson's ratio | $\nu_R$ | | 0.4 | (Alshibli and Hasan, 2009) |
| Coefficient of Expansion | $\alpha_R$ | 1/K | 2.4*10$^{-4}$ | (Agar et al., 2006) |
| Thermal Contact Boundary Layer | $\delta$ | m | 10$^{-6}$ | (Molaro et al., 2015) |

The properties of the regolith are defined following the model of Vasavada et al. (Vasavada et al., 2012; 1999). Unlike basalt, the density and thermal conductivity of regolith are both depth and/or temperature dependent. The regolith density ($\rho_R$) is given by:

$$\rho_R = \rho_d - (\rho_d - \rho_s)\exp\left(\frac{-z}{0.06}\right) \tag{2}$$

where $\rho_s$ is the density at the surface and $\rho_d$ is the density at depth. The regolith thermal conductivity ($k_R$) is given by:

$$k_R = k_d - (k_d - k_s)\exp\left(\frac{-z}{0.06}\right) + \chi k_s \left(\frac{T}{350}\right)^3 \tag{3}$$

where $k_s$ is the solid conductivity at the surface, $k_d$ is the solid conductivity at depth, and $\chi$ is the ratio of the radiative to solid component of $k_R$ at $T = 350$ K. This model has been widely used throughout the literature (Moores, 2016), and is consistent with spacecraft observations.

In order to evaluate the temperature within a boulder throughout the solar day, COMSOL solves the localized heat balance equation (4) for heat transfer in solids over time using an implicit solver and dynamic time step. This is given by:

$$c_p\rho\left(\frac{dT}{dt} + u \cdot \nabla T\right) + \nabla \cdot Q = -\alpha T : \left(\frac{dT}{dt} + u_{trans} \cdot \nabla S\right) \tag{4}$$

where $Q$ is the conductive and radiative heat flux, $u_{trans}$ is the velocity vector of translational motion, $\alpha$ is the coefficient of thermal expansion, and $S$ is the second Piola-

Kirchhoff stress tensor. The (:) operator is the colon, or double dot, product. The right-hand side of equation 4 accounts for thermoelastic damping. The displacement ($u$) is given in terms of the Cauchy stress tensor ($s$):

$$\rho \frac{\partial^2 u}{\partial t^2} = f - \nabla \cdot s \tag{5}$$

where $f$ is the volume force vector and the density is that of the actual deformed state. The stress tensor (s) is related to both the elastic strain ($\varepsilon_{el}$) and inelastic, or in this case thermal, strain tensors ($\varepsilon_{th}$):

$$s = D : (\varepsilon_{el} - \varepsilon_{th}) = D : (\varepsilon_{el} - \alpha(T - T_o)) \tag{6}$$

where $T_o$ is the strain reference temperature, and $D$ is a 4$^{th}$ order elasticity tensor that is a function of the Young's modulus ($E$) and Poisson's ratio ($v$) of the material:

$$D = \frac{E}{(1+v)(1-2v)} \begin{bmatrix} 1-v & v & v & 0 & 0 & 0 \\ v & 1-v & v & 0 & 0 & 0 \\ v & v & 1-v & 0 & 0 & 0 \\ 0 & 0 & 0 & \frac{1-2v}{v} & 0 & 0 \\ 0 & 0 & 0 & 0 & \frac{1-2v}{v} & 0 \\ 0 & 0 & 0 & 0 & 0 & \frac{1-2v}{v} \end{bmatrix} \tag{7}$$

The surface of the model geometry (i.e. the boundary which is open to space) is defined as the top boundary of the regolith volume, and the boundary of the sphere that protrudes above that volume. These are the only boundaries that receive incident radiation and participate in surface-to-surface radiation. Incident solar radiation is applied to the surface by defining the radiation source as a blackbody at infinite distance, with a flux of 1361 W/m$^2$ at 1 AU. The time dependent solar position at the lunar surface (Latitude 0, Longitude 0) is computed separately using the NAIF SPICE Toolkit and supplied to COMSOL via a text file. COMSOL calculates the incident radiation for each point on the surface using this position, accounting for local surface slope and aspect angle of each mesh element. When any part of the surface experiences a local sunrise or sunset, the incident flux is scaled linearly with the portion of the solar disc visible above the local horizon. Accounting for the size of the solar disc is important, as simulating the sun as a point source leads to artificially high stresses in some cases (see Appendix A.2). The ambient temperature above the surface is set to 2 K, to most closely approximate an airless environment.

    The effects of radiative transfer between the rock and regolith surfaces is accounted for by calculating the energy transfer between each mesh element and other elements within its view factor. This is controlled by material absorptivity (one minus the albedo) and emissivity (Tables 1 and 2). COMSOL assumes Kirchoff's Law, and thus the absorptivity and emissivity for a given material are equal in value. We chose an emissivity value of 0.9 and 0.95 for rock and regolith, respectively (Bandfield et al., 2011; Vasavada et al., 2012). To illustrate the robustness of our temperature results, a

model run was conducted on a cubic volume of regolith several meters wide and 4.5 m in depth, with material properties as previously described. The average surface temperature of the regolith is shown in Figure 2. This result agrees remarkably well with Vasavada et al. (2012) (their Figure 9), who selected best-fit parameters for their model by comparing their results to Diviner measurements of regolith temperatures, suggesting that deviations from Kirchoff's law do not produce significant effects

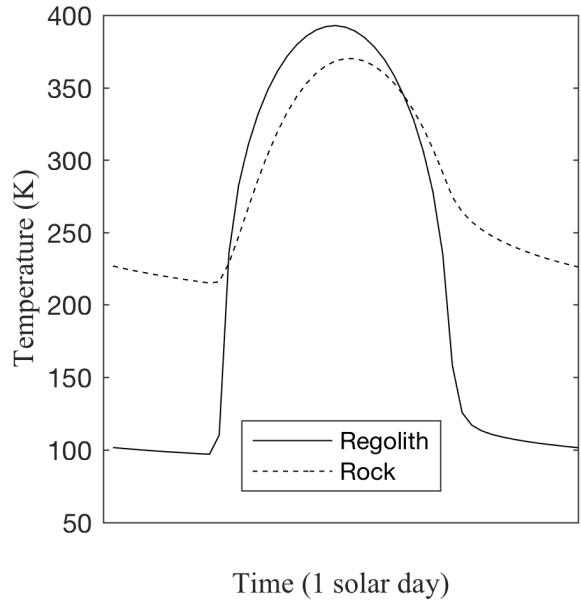

Figure 2. Profiles of the average surface temperature of a block of regolith (solid) and a block of basalt (dotted) with the same dimensions as in Figure 1, but no boulder.

For lunar cases, the initial temperature of boulders ≤ 1 m is set at 280 K. Since these boulders are smaller than the diurnal skin depth, their temperatures normalize quickly over only a few diurnal cycles. For boulders > 1 m, the thermal portion of the calculation was run for as many diurnal cycles as required to stabilize the temperature at the boulder's center. The average temperature of the boulder center was then used as the initial value for the full simulation, typically between 260-270 K depending on boulder size. In all cases, each full simulation was run over several diurnal cycles to ensure that the temperatures and stresses exhibited repeatable (i.e. stable) behavior. The initial temperature of the regolith is 240 K, which is the mean temperature of the regolith at depth at the equator (Vasavada et al., 2012). For model runs with varying solar distance and day length, new initial temperatures of both the boulder and the regolith were determined in a similar way.

The surface is defined as mechanically free, with a strain reference temperature set to each boulder's initial temperature. The sides of the regolith volume are defined as periodic with respect to temperature, and fixed with respect to displacement. The latter was done to improve computational efficiency, and has a negligible effect (see Appendix A.1) on the results due to the regolith's low value of Young's modulus and stable temperature at shallow depths. The bottom boundary of the regolith volume is also held fixed.

Physically, the boulder and regolith volume are discrete objects with different properties. However, running the model with two discrete finite element meshes is computationally expensive. Instead, the objects share a single mesh, and a "Thermal Contact" is defined between them, acting as a boundary layer between domains with differing material properties. The heat transfer across the Thermal Contact boundary is:

$$Q = \frac{k_{eff}}{\delta} \Delta T \quad (8)$$

where $\delta$ is the thickness of the boundary layer. $k_{eff}$ is the effective thermal conductivity, given by:

$$k_{eff} = \frac{2k_R k_B}{(k_R + k_B)} \quad (9)$$

where $k_R$ and $k_B$ are the thermal conductivities of the regolith and the basalt, respectively (Tables 1 and 2). The thickness of the boundary layer is set to $10^{-6}$ m, comparable to a typical grain size for basaltic rock. This value is somewhat arbitrary, and has a negligible effect on results due to the small change in temperature across that boundary (see Appendix A.1).

COMSOL allows the thermal contact boundary layer to move as the objects expand and contract, though the two domains remain connected to each other. Because the temperature does not jump significantly across this boundary in our model, this option provides a reasonable solution that is computationally efficient. In some model runs, this does generate artificial stresses along the contact boundary, which are ~0.3 MPa or smaller. These are reflected in our uncertainty values (see Appendix A.1). Additionally, we present our results in terms of the maximum stress induced within a boulder throughout the day. In all cases, these artificial stresses are always negligible compared to maximum stresses. In the smallest boulders we tested, the physical maximum stresses and artificial stresses become comparable. For this reason, we did not model boulders smaller than 30 cm.

The model calculations can be performed using either a fully coupled or segregated method. A fully coupled model calculates both the heat and displacement equations together for each time step, providing the most accurate calculation. We use the segregated approach, where the result of the thermal calculation is then used as an input for the displacement calculation. This approach can lead to significant inaccuracies for problems with strongly linked physical processes or that are strongly non-linear (e.g., fluid flow). In this study, however, it will produce nearly identical results while being computationally less expensive because the amount of heat generated by the thermal expansion is negligible compared to heating from the sun.

We will present our results in terms of the maximum principal stress ($\sigma_1$), under the convention that tensile stresses are denoted as positive, and compressive stresses as negative ($\sigma_1 > \sigma_2 > \sigma_3$). Thus, the maximum principal stress represents the most amount (if any) of tensile stress available for microcrack propagation at a given time and location. We will use the Maximum Stress Theory as a failure criterion, which states that failure will occur when the maximum principal stress reaches a value equal to some yield stress, determined by failure of a specimen during a simple tension test. Since the maximum principal stress lies along a single plane, comparing it to a simple tension failure criterion is appropriate, as long as the material has an isotropic strength. For rocky materials, this can be considered reasonably true. This failure criterion is commonly used in cases of brittle failure where stress states are simple (Dong et al., 2000; Gustafsson, 1985; McDiarmid, 1987; Ugural and Fenster, 2003). For simplicity, we will refer to the maximum principal stress ($\sigma_1$) as stress ($\sigma$) throughout the rest of the paper.

Rock strengths are on the order of ~100 MPa, though typically only ~25% (or less, depending on the environment) of this is needed to generate fatigue on Earth (Attewell and Farmer, 1974). However, the environment plays a strong role in material strength, and crack propagation in vacuum is not well understood. This makes it challenging to determine a realistic threshold for initiating fatigue on airless body surfaces, as is discussed further in section 4.2.

## 3. Results

### 3.1 Stress in a 1 m Boulder

As the simulated temperature of a boulder changes throughout the day, we can observe its interior on a plane defined by the path of the sun overhead. Figure 3 shows a temperature within a 1 m boulder at two-hour intervals (~2.5 Earth days, for the Moon), beginning at midnight. When the sun rises (panel d), the right side of the boulder begins to heat very quickly. The temperature gradient in (d) is very strong, but declines throughout the following panels (e-h) as heat is conducted to the boulder's interior. In the afternoon, the right side of the boulder moves into shadow and begins to cool (g-i). Following sunset, the entire upper hemisphere is cooling (j), which continues throughout the night. The lower hemisphere takes longer to cool because the surrounding regolith is strongly insulating. The boulder reaches a maximum temperature of ~387 K, and a minimum of ~152 K.

Figure 4 shows the maximum principal stress in the same plane, at the same intervals. Stresses during the night leading up to sunrise (d) are generally low (< 1MPa). During sunrise, higher stresses are induced as the boulder begins to heat up and a temperature gradient is set up within the interior, inducing the stress field shown in panel (d). The right side of the boulder is in compression at the outer edge, and as that edge expands away from the boulder interior it creates a region of strong tension with a peak value of 7 MPa. The magnitude of the induced stress is not related to the value of the thermal gradient itself, but rather to the difference in temperature between the surface and the resulting interior region of tension (i.e., the $\Delta T$ part of $\Delta T/\Delta z$).

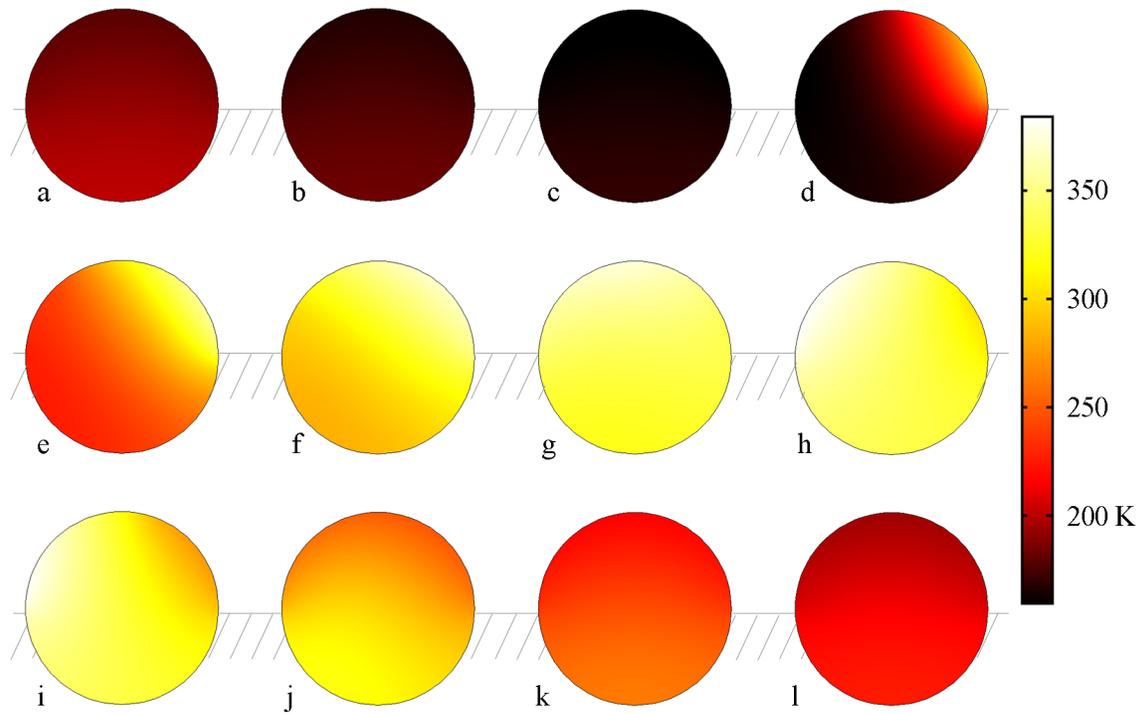

Figure 3. Snapshots of the temperature on a 2D plane through a 1 m boulder (see Figure 1). The cut plane is E-W in orientation, and so the sun moves from right to left across the page. The snapshots begin at an hour angle of 0, and are taken at 2 hour (30°) intervals.

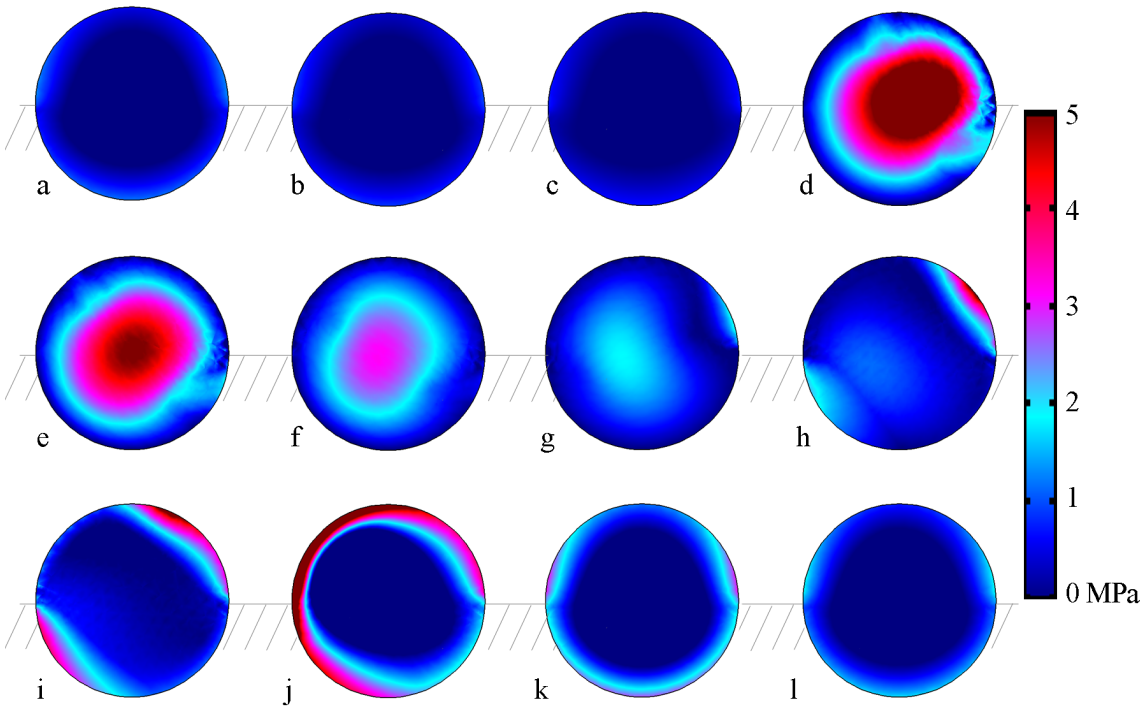

Figure 4. Snapshots of the maximum principal stress on a 2D plane through a 1 m boulder (see Figure 1), corresponding to the temperature snapshots in Figure CC. The snapshots begin at an hour angle of 0 and are taken in 2 hour (30°) intervals, with the exception that (j) shows 17:30 rather than 18 in order to catch the time of peak stress. The peak interior stress is 6.9 MPa (d) and the peak surface stress is 13.2 MPa (j).

In panel (g) of Figure 4, a new region of tension forms on the right side of the boulder as this area moves into afternoon shadow. This region grows as more of the boulder begins to cool until the entire surface moves into a state of tension, peaking after sunset in panel (j) at 13 MPa. In spite of the fact that this stress is higher than that induced at sunrise, the associated macroscopic temperature gradient is small. Unlike the interior stresses, these are surface-parallel stresses at the boulder's exterior caused by the contraction of the surface. Thus we have two distinct stressing mechanisms. At the boulder's surface, the combination of thermal contraction and a free mechanical surface produces high stresses even in the absence of thermal gradients between the surface and the interior. In the interior, the macroscopic thermal gradient sets up a stress field where the magnitude of induced stress is controlled by the difference in temperature between the surface and the resulting interior region of tension. These mechanisms dominate at different times of day at different locations within the rock and cannot always be easily separated.

We can also observe the stress response in the boulder due to the radiative interaction with the regolith. Most notably, in panels (k) through (c) of Figure 4, it takes longer for the stresses on the exposed sides of the boulder to dissipate, as they cool less efficiently. This also occurs at the boulder's bottom edge due to insulation from the regolith.

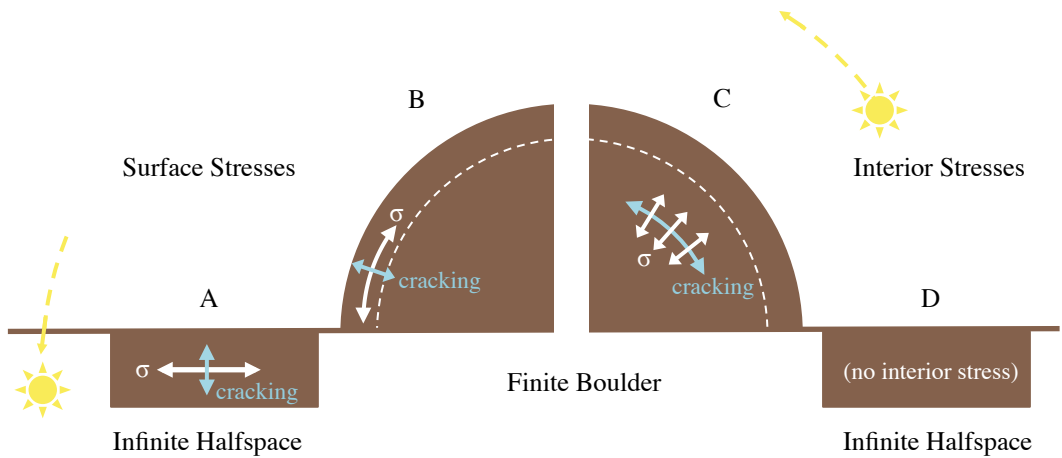

Figure 5. View of the locations of surface (A and B) and interior (C and D) stresses in an infinite halfspace and a finite boulder. Surface stresses occur during daytime heating, and interior stresses occur after sunset. The white arrows denote the orientation of the maximum principal stress, and the red arrows denote the resulting direction of crack propagation.

3.2 Effects of Boulder Diameter

We can investigate the role that size plays in the bimodal response of a boulder to diurnal stress cycles by discussing interior and surface stresses separately. After sunrise, the boulder's surface is in compression and the interior is in tension. Between these two regions exists a plane where the stress is zero. We refer to interior stresses as those interior to this neutral plane (Figure 5, right). After sunset, the boulder's surface is in tension and the interior is in compression. Similarly, we refer to surface stresses as those in the region exterior to the neutral plane (Figure 5 left). We will consider how each of these stress mechanisms evolves independently with boulder size.

In order to best understand our results, consider two theoretical end member cases. Interior stresses are controlled by a temperature gradient set up within boulders of finite size after sunrise (Figure 5, right). As boulders become larger, their behavior will eventually approach that of an infinite halfspace (Figure 5, a), where no interior stresses are present. An infinite rock undergoing compression at the surface does still have tensile stresses at depth as the thermal wave travels downward, however since these have the same effect as surface stresses (see section 4.1) we do not include them in our discussion of interior stresses. Interior stresses are shown for boulders up to 20 m.

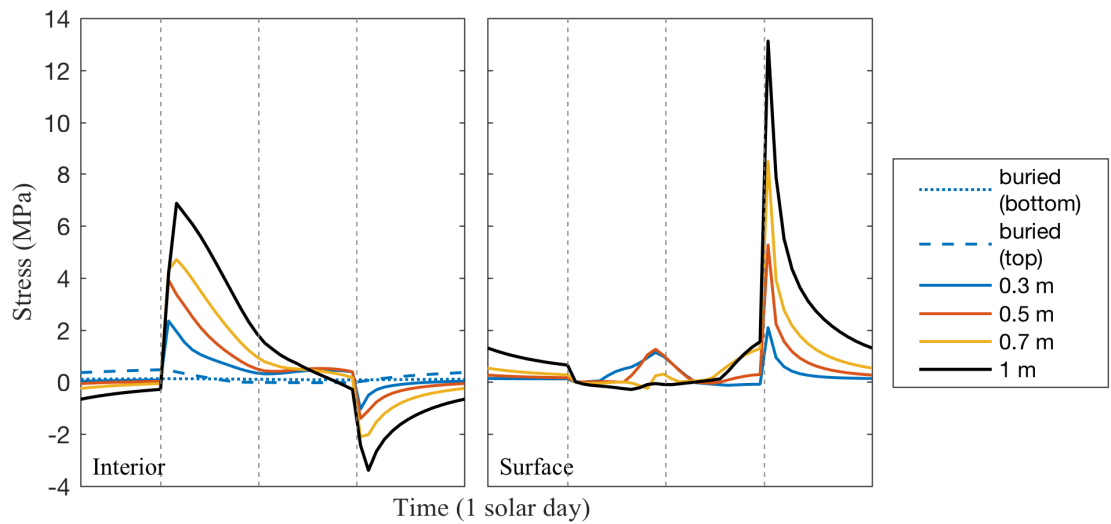

Figure 6. Profiles of the interior (left) and surface (right) stresses in boulders with diameters from 0.3 to 1 m throughout the lunar day, taken at the location in the boulder where the maximum interior or surface stress occurs. The dotted lines show the times of sunrise, noon, and sunset.

Figure 6 (left) shows the evolution of the peak interior stresses over a diurnal cycle in boulders with diameters from 0.3 to 1 m. Each profile represents the stress throughout the course of the day at the location in the boulder where the maximum interior stress occurs. These profiles show that peak stress is a strong function of size, with smaller peak stresses induced in smaller boulders. The peak interior stress experienced by a 1 m boulder is ~7 MPa, while a 0.3 m boulder peaks at only ~2 MPa.

Boulders smaller than the diurnal skin depth (~0.8 m) are able to cool efficiently and thermally equilibrate quickly. In most boulders, interior stresses are weaker than surface stresses (Figure 6, right). However, the 0.3 m boulder displays the opposite behavior. Combined with its overall weak response to the diurnal cycle, this suggests it may represent a size threshold below which thermally induced breakdown occurs very slowly or not at all. This is discussed further in the following sections.

As boulders increase from 1 m to 6 m (Figure 7, top left), the interior stresses increase and peak values occur at later times of day as the thermal wave penetrates deeper into the rock, setting up the temperature gradient over a longer period of time. This is reflected in the shift in location of peak interior stresses as boulder size increases (Figure 8). For boulders ≥10 m (Figure 7, bottom left), interior stresses begin to decrease as the thermal wave loses contact with the boulder center. The lower hemisphere of boulders ≥10 m comes into near thermal equilibrium with the regolith, and thus stresses are primarily limited to their top hemispheres.

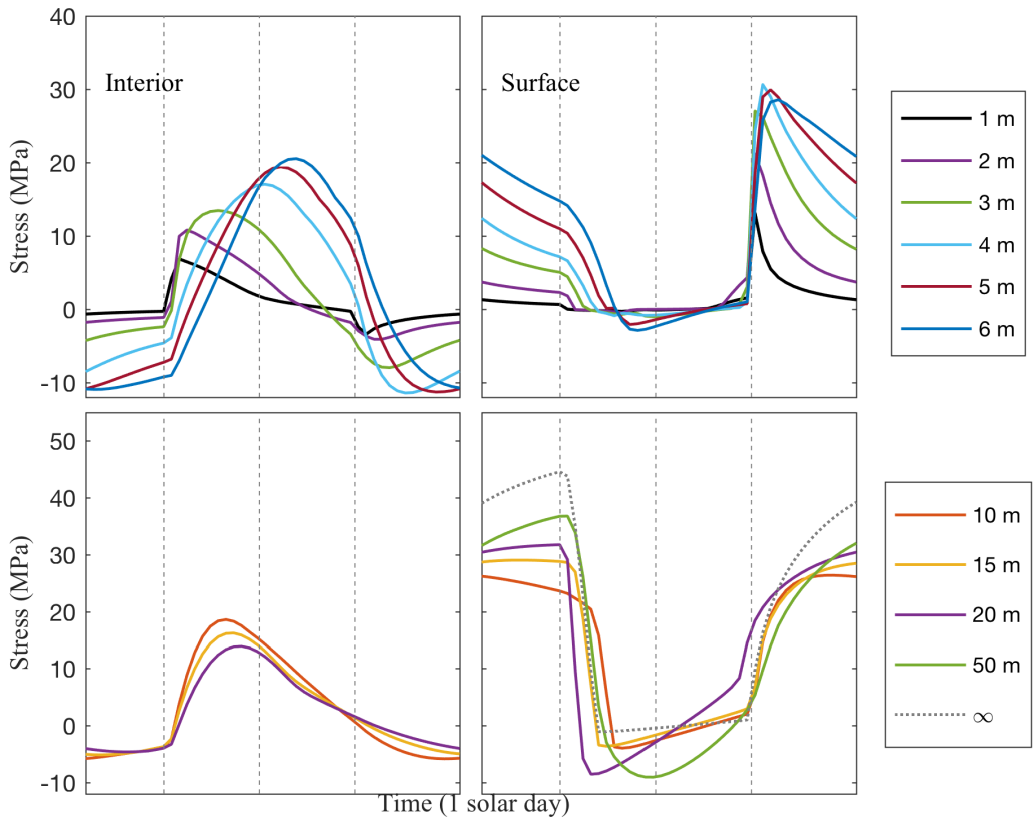

Figure 7. Profiles of the interior (left) and surface (right) stresses in boulders throughout the lunar day, taken at the location in the boulder where the maximum interior or surface stress occurs. The top row shows boulders with diameters from 1 m to 6 m, and the bottom row from 10 m to 50 m. There is no 50 m profile in the lower left panel, as interior stresses were only calculated for boulders up to 20 m. The dotted profile in the lower right panel is the stress at the surface of a solid block of basalt, as modeled in Figure 2, representing an infinite halfspace. The vertical dotted lines show the times of sunrise, noon, and sunset.

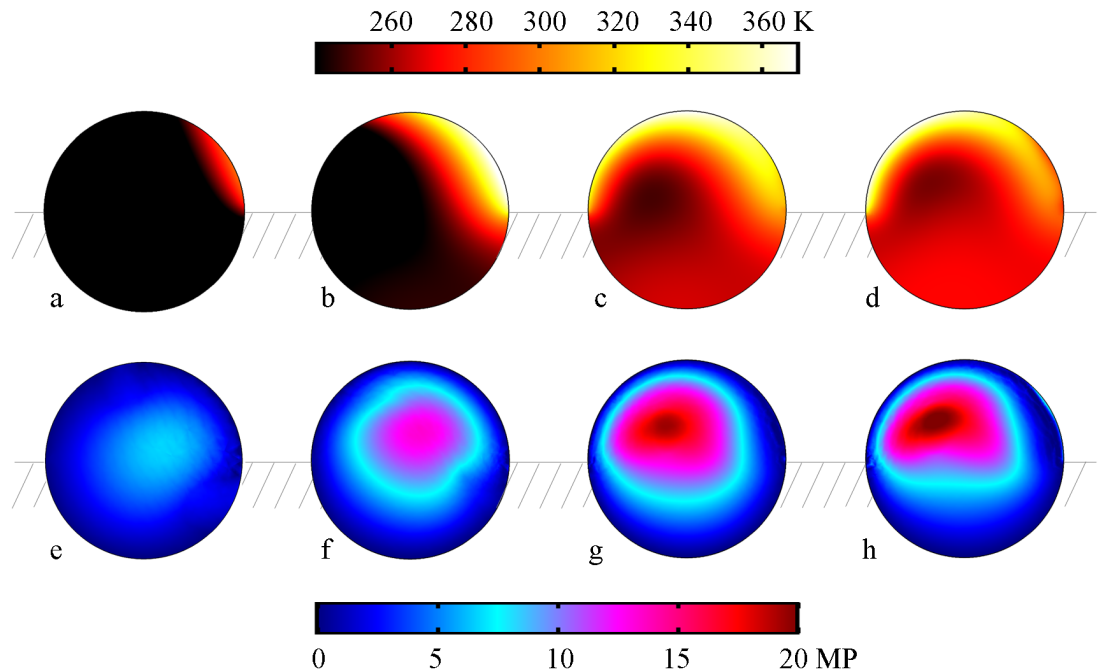

Figure 8. Snapshots of the temperature (top) and stress (bottom) at the time of peak interior stress, on a 2D plane (see Figure 1) through a 1 m (a, e), 3 m (b, f), 5 m (c, g), and 7 m (d, h) boulder. The snapshots take place at 6:30 am, 9 am, 1 pm, and 11:30 am, respectively.

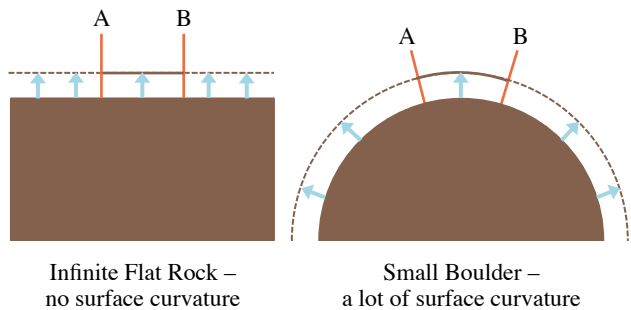

Figure 9. For a large boulder (left) with little to surface curvature, no increase in the distance between points A and B occurs during expansion. For a small boulder (right), however, the surface curvature causes an increase in circumference during expansion, moving points A and B farther apart and relieving stress.

Surface stresses occur as a result of cooling and contraction of the boulder surface (Figure 5, a, b). To understand how they evolve with boulder size, we again consider our end member cases. An infinite flat rock subjected to a thermal cycle will expand during heating and contract during cooling. Since the rock is infinite, an element at its surface will not experience a change in its surface area as this occurs (Figure 9, left), and the induced stress will be determined by the amount of expansion or contraction it experiences. This amount is fixed because it is limited to the volume within approximately one skin depth of the surface. However, a finite rock undergoing expansion or contraction will experience a change in its surface area, which will relieve some stress that would otherwise be induced (Figure 9, right). In our study, smaller boulders have the highest surface curvature and thus will experience the most stress relief. Larger boulders will approach the infinite halfspace scenario where no stress can be relieved during expansion or contraction. On the other hand, surface curvature also contributes to increased peak stress in small boulders for a different reason. Their large surface area to volume ratio allows them to cool more efficiently than an infinite halfspace. For very small boulders, this causes peak surface stresses to occur during and just after sunset when these heat fluxes cause strong surface contraction. Thus, peak surface stresses are controlled by two effects, efficient cooling and decreased surface curvature, in boulders of different sizes.

Peak surface stresses for boulders between 0.3 m and 1 m are shown in Figure 6, (right), and in Figure 7 (top right) for boulders between 1 m and 6 m. These stresses occur during and just after sunset, and are associated with strong radiative cooling that remains nearly constant with diameter for boulders ≤ 3 m. At larger diameters, the heat flux at sunset declines rapidly, as boulders cool less efficiently due to their larger internal heat reservoir. The increase in stress at sunset with boulder size reverses for boulders > 4 m.

Boulders ≥ 10 m (Figure 7, bottom right) experience continuous cooling and contraction throughout the night, so peak surface stresses occur just before sunrise. These stresses increase due to the decrease in surface curvature with increasing diameter. The trend in these profiles is somewhat less smooth than in other panels of Figure 7 because the location of peak stress, and time of day it occurs, varies more significantly in larger boulders.

Figure 10 shows peak stresses for boulders of all diameters. Peak surface stresses in small boulders are limited by their cooling efficiency at sunset, whereas those in large boulders are controlled by their surface curvature. Boulders between 3 m and 7 m experience overlap between these two effects. Stresses in this range peak at 4 m (~5x the diurnal skin depth) where a local maximum along the stress-diameter curve is observed (Figure 10). For boulders ≥ 10 m, the data can be fit with an exponential curve using a weighted least squares method, where the weight of each point is one over the uncertainty (see Appendix A.1) squared. The curve is given by:

$$\sigma_{D \geq 10 \text{ m}} = 45.5\exp(-6.628 * 10^{-6} D) - 26.2\exp(-0.03131 D) \qquad (10)$$

Points without visible error bars have an uncertainty of $\pm$ 0.5 MPa. The value for peak stress in an infinite halfspace is $45 \pm 2$ MPa. Despite the uncertainties in the largest boulders, this curves gives a reasonable fit with an $R^2$ value of 0.9998.

While we do not consider their effects in this study, peak compressional stresses during the day also change with boulder size. Since tensile stresses are always positive, analyzing compressional stresses would be best done by plotting the minimum principle stress, rather than the maximum as we present. In Figure 7, the times at which a profile is negative indicates that none of the principal stresses are tensile, and the plotted value represents the weakest compressional stress component at that location. While these compressional stresses do follow a trend with boulder diameter, they do not provide meaningful insight.

We also investigate the effect of regolith cover on induced stresses by simulating a 0.3 m boulder completely buried such that its top boundary is 2.5 cm below the surface of the regolith. The dashed and dotted lines in Figure 6 (left) show the stress at its top (dashed) and bottom (dotted) edges. The regolith provides such strong insulation that the boulder's diurnal temperature range is reduced to a mere ~13 K, resulting in peak stresses of 0.47 and 0.12 MPa. These are comparable to the artificial stresses induced at the thermal contact boundary (average ~0.3 MPa), indicating that the diurnal temperature cycle is unable to induce significant stresses within buried boulders, effectively shielding them from the fatigue process.

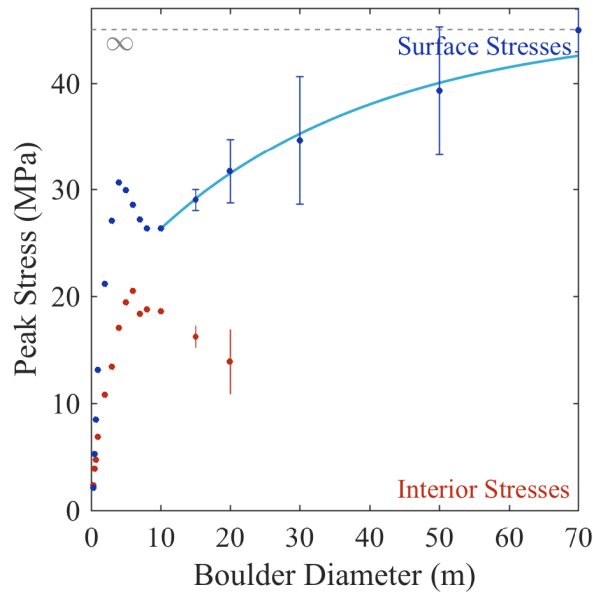

Figure 10. Peak surface (blue, top) and interior (red, bottom) stresses experienced in boulders of varying diameter. Points without visible error bars have an uncertainty of 0.5 MPa. The curve represents the best fit to the surface stresses for boulders ≥10 m. The dotted line is the stress at the surface of a solid block of basalt, as modeled in Figure 2, representing an infinite halfspace.

3.3 Rock Lithology

While basalt is a common rock type found on the lunar mare, the highlands are anorthositic in composition. The properties of geologic materials do vary, however they are generally on the same order of magnitude as the values used for basalt in this study. Precise values for each property are highly unique to individual rocks and rock types, making it challenging to model different rock types without testing actual samples. In order to investigate the role each parameter plays in thermally induced stresses, we run test simulations in which all rock properties remain the same as in Table 2, with the exception of a single parameter. By changing one parameter at a time, we can investigate the effect individual material properties have on induced stresses. The test values for each property are shown in Table 3, and represent reasonable variation for geologic materials. The density, emissivity, and heat capacity only exert a weak influence over induced interior and surface stresses. Lowering the thermal conductivity has a moderate influence, increasing interior stresses by 3.5 MPa due by increasing the temperature gradients, and decreasing surface stresses by 1.6 MPa due to the lowered heat flux during cooling.

As is the case in on mineral-grain scales (Molaro et al., 2015), the coefficient of thermal expansion and Young's modulus have the strongest effect on thermally induced stresses in boulders. An increase in the Young's modulus results in an increase in stress of 13 MPa and 7 MPa for interior and surface stresses, respectively. An increase in the thermal expansion coefficient also yields higher stresses, with an increase of 26 MPa and 14 MPa, respectively. In geologic materials, the typical range for Young's modulus is 10-100 MPa, and for the coefficient of thermal expansion 1-3 $10^{-5}$ $K^{-1}$. Thus, the parameters used in this study yield somewhat of a lower limit on induced stress values, and inform the qualitative relationship between boulders of different size. When considering other rock types, Table 3 gives a general sense for how stresses will change with rock type. For example, the lunar highlands are largely composed of anorthosite which has a Young's modulus of ~100 GPa (Turcotte and Schubert, 2014), suggesting that highlands material experience stresses a factor of 2 higher than basalt, and so may breakdown faster.

Ultimately, this sensitivity test shows that thermally induced stresses will be unique to different planetary materials in different thermal environments, and expected behavior and lifetimes may require unique calculations. Differences in induced stress with varying properties is also likely to be exaggerated in environments with larger temperature ranges, making this effect of more or less significance on different solar system bodies. Other factors may also add complications, such as anisotropy in a material's thermal expansion coefficient, which plays a significant role in weathering of granite and marble on Earth.

| | Parameter | Standard | Test | $\Delta\sigma$ (MPa) | $\Delta\sigma$ (%) | Trend | Strength |
|---|---|---|---|---|---|---|---|
| Interior Stress | $\rho_B$ (kg/m³) | 3150 | 3000 | +0.09 | 1.3 | ↓ $\rho$  ↑ $\sigma$ | Weak |
| | $c_p$ (J/kg K) | $c_p$ | $c_p * 1.05$ | +0.19 | 2.7 | ↑ $c_p$  ↑ $\sigma$ | Weak |
| | $\varepsilon_B$ | 0.9 | 0.8 | -0.96 | -14 | ↓ $\epsilon$  ↓ $\sigma$ | Weak |
| | $k_B$ (W/m K) | 2 | 1.5 | +3.5 | 50 | ↓ $k$  ↑ $\sigma$ | Moderate |
| | $E_B$ (GPa) | 50 | 100 | +13 | 186 | ↑ $E$  ↑ $\sigma$ | Strong |
| | $\alpha_B$ (1/K) | $1*10^{-5}$ | $3*10^{-5}$ | +26 | 370 | ↑ $\alpha$ ↑ $\sigma$ | Strong |
| Surface Stress | $\rho_B$ (kg/m³) | 3150 | 3000 | -0.04 | -0.3 | ↓ $\rho$  ↓ $\sigma$ | Weak |
| | $c_p$ (J/kg K) | $c_p$ | $c_p * 1.05$ | -0.04 | -0.3 | ↑ $c_p$  ↓ $\sigma$ | Weak |
| | $\varepsilon_B$ | 0.9 | 0.8 | -0.55 | -4.2 | ↓ $\epsilon$  ↓ $\sigma$ | Weak |
| | $k_B$ (W/m K) | 2 | 1.5 | -1.6 | -12 | ↓ $k$  ↓ $\sigma$ | Moderate |
| | $E_B$ (GPa) | 50 | 100 | +6.9 | 53 | ↑ $E$  ↑ $\sigma$ | Strong |
| | $\alpha_B$ (1/K) | $1*10^{-5}$ | $3*10^{-5}$ | +14 | 108 | ↑ $\alpha$ ↑ $\sigma$ | Strong |

3.4 Stresses on Other Bodies

Other bodies in the inner solar system provide a range of thermal environments in which we can explore this process. To do so, we calculate the incident solar flux for an arbitrary surface with varying solar distance and day length. These objects are assumed to have zero obliquity and eccentricity. Figure 11 is a contour plot showing peak stresses in a 1 m boulder on these arbitrary surfaces, which range from ~1-70 MPa. As expected (Molaro and Byrne, 2012; Molaro et al., 2015), objects that rotate slowly and/or that are close to the sun experience the highest stresses. Solar distance plays the primary role, as peak stresses occur from surface contraction at sunset and thus are controlled by the maximum temperature achieved at the boulder's surface. Which mechanism is dominant on different bodies may change depending on the type of stress (interior or surface), boulder size, and time of day of interest. For a 1 m boulder with a 0.1 Earth day rotation period, peak stress follows the relationship $\sigma = 4.98s^{-1.65}$ where s is the solar distance. The value of the exponent varies slightly with boulder diameter, with an average of $-1.71 \pm 0.1$ for a 30 cm, 1m, and 10 m boulder at this rotation rate. Solar distance has less influence on stress for very slowly rotating bodies, for example the exponent for a 1 m boulder decreases from -1.5 to -1.2, for a rotation period of 1 and 5 Earth days, respectively.

While the objects we modeled were arbitrary, the perihelion positions of some notable solar system bodies are marked in Figure 11 for reference. The results suggest that objects like Mercury and 3200 Phaethon may be significantly affected by thermal breakdown, with stresses estimated to exceed 70 MPa and 50 MPa, respectively. In contrast, objects such as Itokawa, Eros, and Vesta have relatively low stresses, suggesting thermal breakdown may be inactive on their surfaces. On the other hand, if even a small amount of crack growth were to occur each day, breakdown could happen quickly due to their high cycling rate. Ultimately, however, we do not yet know the efficacy of this process on these surfaces, as there is significant uncertainty in our understanding of the stress required to drive crack propagation on airless bodies (see section 4.2).

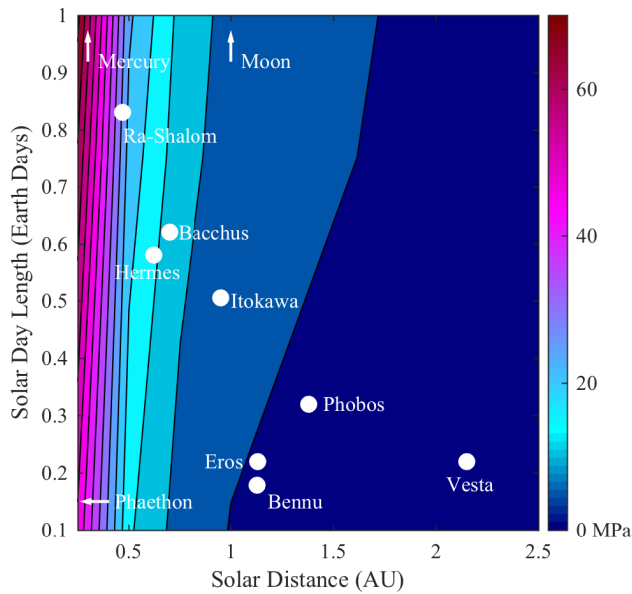

Figure 11. Contour plot of peak stresses in 1 m boulders in regolith, as with the previous cases, on bodies with arbitrary solar distances and day lengths. The locations of particular objects of interest are noted on the plot at their perihelion locations, though they were not specifically modeled.

It is important to note that we use the same regolith and rock properties and geometry as for the lunar cases, thus these results show stresses induced on "lunar-like" surfaces. This provides a general sense of relative stresses throughout the inner solar system. In reality, the surfaces of asteroids will vary substantially in composition, regolith depth, and boulder size and distribution, all of which will affect induced stresses. For example, studies of meteorite properties (e.g., Opiel et al., 2010, 2012; Cotto-Figueroa et al., 2012) have found that the thermal conductivity of CM chondrites is ~0.5 W/m K and the Young's modulus of CV chondrites is on the order of 10 MPa, suggesting (per Table 3) that the stresses in Figure 11 could vary by a factor of ~2. Additionally, the boulder and regolith thermally and radiatively interact with each other, suggesting that asteroids with little to no know regolith cover, or those that are rubble piles, will respond differently to thermal forcing. Overall, the range of surface characteristics and orbital properties seen throughout the asteroid population suggests a range of diverse and unique responses to thermal cycling, emphasizing the need for additional research on this topic. Future work will focus more specifically on understanding thermally induced breakdown on asteroid surfaces.

4. Damage Modeling

4.1 Direction of Crack Propagation

   The orientation of stresses within boulders can provide insight into how breakdown may occur. Microcracks will primarily propagate perpendicular to the direction of tensile stress, assuming it is large enough to overcome material tensile strength. For the interior and surface stresses we discuss here, the maximum principal stress (as presented in the above results) dominates the stress field at the times and locations within the boulders they occur (i.e., each of the other two principal stresses either are negligible or act in the same plane as the maximum principal stress). Thus the maximum principal stress will effectively control the plane in which the direction of crack propagation occurs.
   The interior and surface stresses described in the previous section result from different mechanisms, and thus have different orientations. Interior stresses act on a plane perpendicular to the path of the sun (Figure 5), driving the propagation of surface-parallel cracks and contributing to the exfoliation of planar fragments. In contrast, surface stresses act parallel to the boulder surface. These drive the propagation of surface-perpendicular cracks, contributing to granular disintegration. These two mechanisms likely work together to hasten disaggregation of near-surface material.

4.2 Estimated Boulder Lifetimes

While interest in thermally induced rock breakdown is growing, relating the results from models (Molaro et al., 2015) and experiments (Delbo et al., 2014) to estimates of breakdown and regolith production rates presents a unique challenge. Most of the literature studying thermal fatigue focuses on engineering materials and applications. Limited work has been done on developing damage laws for geologic materials, let alone for objects in vacuum and/or over large temperature ranges.
       The most commonly used fatigue model is the Paris law, which describes the kinetics of crack growth under cyclic loading (Paris and Erdogan, 1963). The Paris law has been adopted to model the fatigue behavior of geological materials, such as sandstones (Le et al., 2014; Migliazza et al., 2011), and has recently been used by Delbo et al. (2014) to estimate the lifetimes of rocks on asteroids. There are many complications to implementing the Paris law in a planetary context, which will be described below. First, we will demonstrate an example and compare our results to those of Delbo et al. (2014).
       The Paris law relates the crack growth per cycle ($da/dN$) to the amplitude of the applied stress intensity factor ($\Delta K$):

$$\frac{da}{dN} = C\Delta K^n \tag{11}$$

where $a$ is the crack length, $N$ is the number of cycles, and $C$ and $n$ are empirically determined parameters. Following Delbo et al. (2014), we use values for $C$ and $n$ of 0.0003 $(m/(MPa \sqrt{m})^{3.84})$ and 3.84, respectively, which were determined by Migliazza et al. (2011) for marble. While marble is an imperfect match for planetary materials, particularly given its large grain size and anisotropic coefficient of expansion, few studies have been done determining these parameters for geologic materials.

The stress intensity factor is a function of the stress amplitude of the applied load ($\Delta\sigma$). In our case, the nature of the thermally induced stress field is spatiotemporally complex (e.g., Figure 4), and we will express the stress amplitude as $\Delta\sigma(x,y,z,t) = \Delta\sigma_0 h(x,y,z,t)$, where $\Delta\sigma_0$ is the maximum stress experienced by the boulder (Figure 6). For a 30 cm boulder, $\Delta\sigma_0$ is ~1.5 MPa. By dimensional analysis, the amplitude of the applied stress intensity factor is then given by:

$$\Delta K(a) = \Delta\sigma_0 \sqrt{D} f(\alpha) \tag{12}$$

where $D$ is the characteristic size of the boulder, and $\alpha$ is the relative crack size given by the ratio $a/D$. The dimensionless stress intensity factor $f(x)$ is determined by the specimen geometry and the stress field, and not always trivial to define. It must be determined numerically, except for simple loading configurations for which analytical solutions are available. The simplest case is a crack in an infinite medium subject to Mode I loading (i.e., the stress field is perpendicular to the crack) of amplitude $\Delta\sigma_0$, for which $f(\alpha) = \sqrt{\pi\alpha}$. For objects of finite size, with different crack geometries and loading configurations, $f(\alpha)$ can take more complex forms (Tada et al., 2000). In our case, $f(x)$ must reflect the size of the boulder, crack geometry, and the spatiotemporal nature of our stress field given by $h(x,y,z,t)$, and its appropriate form is unknown. For the crack size and geometry considered in this study, the choice of a more exact function for a finite-size body (e.g., page 40, Tada et al., 2000) does not change our result at the order of magnitude level. Given the idealized nature of our boulders, the factor $f(\alpha)$ as defined above will allow us to make a reasonable calculation for the duration of fatigue crack growth.

Per equation 11, a small amount of crack growth occurs as the boulder undergoes each stress cycle. As the crack grows in length, the value of the stress intensity factor ($\Delta K$) at the crack tip increases. When $\Delta K$ reaches the material's fracture toughness ($K_c$), the specimen is no longer able to sustain the prescribed load and the crack will grow in an unstable manner. The critical crack length at which this occurs is denoted ($a_c$). By substituting equation 12 into equation 11 and integrating, we can calculate the total number of cycles ($N_f$) until the crack reaches $a_c$:

$$N_f = C^{-1}\Delta\sigma_0^{-n} D^{-\frac{n}{2}+1} \int_{\alpha_0}^{\alpha_c} \frac{d\alpha}{f(\alpha)^{n/2}} \tag{13}$$

where $\alpha_0$ is the initial relative crack size ($a_0/D$) and $\alpha_c$ is the critical relative crack size ($a_c/D$). Based on the crack geometry considered here, the critical crack length $a_c$ can be estimated as:

$$a_c = \frac{1}{\pi}\left(\frac{K_c}{\sigma_{max}}\right)^2 \qquad (14)$$

Using $\sigma_{max}$ = 2.4 MPa from our results, and $K_c$ = 1.35 MPa*m$^{1/2}$ (Migliazza et al., 2011) we obtain a critical crack length of 10 cm. We assume an initial crack length of 30 $\mu$m, following Delbo et al. (2014). Based on equation 13, we obtain a crack growth duration of ~1.3 x 10$^6$ cycles, or ~10$^5$ years for a 30 cm boulder on the surface of the Moon, which is consistent with the 10$^3$-10$^6$ years calculated by Delbo et al. (2014) for a 10 cm boulder on an asteroid in spite of using a simplified stress intensity factor. However, as we discuss below, there are a number of factors that contribute to a lack of confidence in the accuracy and/or usefulness of these calculations.

While the Paris law provides a straightforward means for calculating the duration of fatigue crack growth, it is worthwhile to point out that this duration is not equal to the *total* fatigue lifetime of the specimen. Fatigue failure typically consists of three stages including 1) crack initiation, 2) stable crack growth, and 3) unstable crack growth. The Paris law primarily describes the crack behavior during stage 2. However, since the duration of unstable crack growth is very short, equation 13 can reasonably quantify the total duration of stages 2) and 3). Stage 1 involves microcrack coalescence and macrocrack initiation, which is not described by the Paris law. This stage can account for the majority (>70%) of an object's total lifetime (Janssen et al., 2002), and thus equation 13 may represent only a small part of the entire breakdown process. On the other hand, if the specimen contains a large crack prior to loading, then stage 1 is essentially absent. In this case, equation 13 would give a reasonable estimate of the total lifetime. In this context, the above calculation is performed inconsistently. Delbo et al. (2014) measured crack growth rates and constrained the values of $C$ and $n$ for a macroscopic crack (~1 mm), and therefore in stage 2 of fatigue crack growth. However, to calculate their sample lifetime, they used an initial crack length of 30 $\mu$m, which falls into the stage 1 regime and is not described by the Paris Law. Thus, their calculation (and ours) significantly underestimates the total lifetime by applying unrealistically high crack growth rates to the early stages of the fatigue process. If we instead repeat our calculation using an initial crack length of 1 mm, it reduces our crack growth duration from 10$^5$ to 10$^3$ years, but does not quantify the length of stage 1.

Furthermore, fatigue crack growth is heavily influenced by the environment. The presence of water vapor, in particular, enhances fatigue crack growth rates and decreases the stress threshold required initiate crack propagation in metals, however similar effects have also been demonstrated in salty and anhydrous environments, including dry air (e.g., Wei 1970; Kirby and Beevers, 1979; Khobaib et al., 1980; Stanzi et al., 1991). As discussed by Janssen et al. (2002), the strength of these environmental effects is influenced by factors such as cycling frequency, loading waveform, and temperature, and overall they note that interactions at crack tips are highly complex and our understanding of the role environment plays is limited. This makes it challenging to quantify uncertainty when applying crack growth parameters from terrestrial studies (e.g., Delbo et al., (2014), done in dry air) to a vacuum environment. Research on crack growth in vacuum is limited for geologic applications, however the increase in stress threshold has been demonstrated in basalt (Krokosky and Husak, 1968). This has the effect of increasing the length of time spent during crack initiation (stage 1), and since this period is not captured in a Paris Law calculation, such estimates of crack growth duration will be even less representative of the object's entire lifetime.

Performing the above calculation for larger boulders leads to further uncertainty in how the Paris Law can be used to understand thermal breakdown on airless bodies. Using the same parameters as described above, the calculation for a 1 m boulder (with $\Delta\sigma_o = 11.9$ MPa) yields a fatigue crack duration of only ~35 years. This clearly seems unrealistic, even in spite of our limited understanding of this process. The shorter crack growth duration results partly from the larger stress amplitude of the 1 m boulder, and thus the calculation will be affected by changes in material properties (section 3.3). However, even a reduction in stress by half yields a fatigue crack duration of only ~500 years, suggesting other factors are at play. In addition to those described above, such a discrepancy may also result from the following effects:

1) The empirical parameters $C$ and $n$ depend on a number of factors, such as chemical environment (as described above), maximum-to-minimum stress ratio, ambient temperature, specimen size, composition, and cycling frequency (Barenblatt and Botvina, 1980; Bazant and Xu, 1991; Ciavarella et al., 2008; Le et al., 2014; Ritchie, 2005). Therefore, even if the Paris Law provides a reasonable estimate for the lifetime a given boulder, it cannot be used to quantify relative lifetimes between boulders of different diameters or in different thermal environments without determining new parameters for each case. The change in diameter itself would only cause a difference in crack growth duration of up to one order of magnitude. However, different factors influence $C$ and $n$ to different extents, and their combined influence can lead to changes anywhere from factors to several orders of magnitude.
2) The present calculation assumes the dimensionless stress intensity factor is the same for both the 30 cm and 1 m boulders. In reality, the spatiotemporal nature of the thermally induced stress field varies with boulder size, which would lead changes in the expression for $f(x)$. However, variations in $f(x)$ alone would not result in a 4-5 order of magnitude change in the crack growth duration. Therefore, it is likely that inaccuracies from both $f(x)$ and $C$ and $n$ are influencing our calculations.

Another approach is to constrain the model parameters by comparing the crack growth duration estimates with observations of lunar rock abundance. Basilevsky et al., (2013) found that no boulders larger than 4 m in diameter were found at craters older than ~150 Ma. To match this age, our calculation for a 5 m boulder ($\Delta\sigma_0 = 30$ MPa) requires a value of $n = 21$. However, in this case our 30 cm and 1 m boulders would never break down due to thermally induced stresses, which is in contradiction with the conclusions of Delbo et al. (2014). Since the efficacy of thermal breakdown relative to micrometeorite bombardment is unknown, this approach only provides an upper limit on $n$, leaving a wide range of implications for intermediate values.

Overall, the empirical parameters and stress intensity factors involved in using the Paris Law cannot easily be generalized from a single experiment to objects with different sizes, crack lengths, or compositions, or those in different thermal and chemical environments, and the Paris Law does not describe all stages of the fatigue process. The effects described above contribute significant uncertainty both to our calculations and those of Delbo et al. (2014). Additionally, the complex spatiotemporal nature of thermally induced stress fields in boulders suggests that a more comprehensive model is needed to describe breakdown through the propagation of many distributed micro- and macroscopic cracks, rather than a single crack as described by the Paris Law. All of this emphasizes the need for more laboratory studies to explore the relationship between thermal cycling crack propagation and distribution, and to measure crack growth parameters in planetary materials and vacuum environments. More data on the size-frequency distribution of rocks on planetary surfaces may also offer a different way of relating models of thermal stress to breakdown rates in view of the size-dependence of fatigue kinetics.

5. Discussion and Conclusions

Our results show that boulders exhibit a bimodal response to diurnal thermal forcing, where interior stresses associated with macroscopic temperature gradients are induced during daytime surface heating, and surface stresses are induced at or after sunset due to cooling and contraction of the surface. The fact that these stresses controlled by multiple mechanisms, and occur at different locations within boulders and at different times of day, emphasizes the complex nature of this problem and highlights the challenges in understanding how breakdown may occur. Nevertheless, we can gain some insight into the differences in the thermal response of boulders of various sizes, and what implications that may have for breakdown.

Peak tensile stresses are on the order of magnitude of 1-10s of MPa, with higher stresses occurring in larger boulders. However, it is important to emphasize that these stresses represent idealized amounts of energy available for crack propagation. It is tempting to suggest that since larger boulders experience higher stresses, they must break down faster. Measurements of rock strengths also indicate that larger boulders typically have lower strengths due to their high volume of pre-existing damage, and structural defects and weaknesses. However, movement of microcrack walls and new crack propagation can also relieve stress, serving to lower the effective elastic modulus. Since the elastic modulus exhibits such a strong control over induced stress, this could lower realistic stresses in very large boulders significantly. While smaller boulders may have less pre-existing damage than larger boulders, any new crack propagation will still relieve some stress, suggesting that, irrespective of size, boulders from new impacts that have not experienced significant thermal processing will be most susceptible to thermal breakdown. Our results also suggest that rocks shielded by even a thin layer of regolith are isolated from any significant thermal stresses. This leads to complications where boulders may develop partial regolith cover via disaggregation and granular disintegration of their surfaces, which may lower the amplitude of their internal stress fields. Additionally, even if the low stresses induced in smaller rocks were to cause them to break down at slower rates, they may disappear faster from burial, which will affect the relative survival times of rocks of each size.

We can further explore the relationship between breakdown and size by considering the location of stresses within boulders. Interior stresses drive the propagation of surface-parallel cracks and contribute to surface exfoliation. Surface stresses drive the propagation of surface-perpendicular cracks, which contribute to granular disintegration. These mechanisms work together to weaken and disaggregate near-surface material. Interior stresses may also contribute to through-going cracks in smaller boulders, however for large boulders, the area affected by thermally induced stresses is limited to the near-surface. Thus, size plays an important consideration beyond simply the amplitude of induced stress, leading us to consider breakdown rates and object lifetimes separately. Larger boulders experience higher stresses than smaller boulders, suggesting they may have faster crack propagation rates. However, they have much more material to erode and less (relative) volume undergoing cyclic stress, suggesting they may have a longer lifetime in spite of a potentially faster breakdown rate. On the other hand, boulders smaller than the diurnal skin depth experience stresses throughout their entire volume, but may have much slower crack propagation rates and thus also have a long lifetime. This leads to interesting questions about how breakdown rates change over time as the boulders shrink, and what that means for the size-frequency distribution expected on the lunar surface. For example, a 4 m boulder experiences only moderately lower stress than those > 10 m, but has significantly less volume, suggesting that it may be preferentially removed by this process. If this were the case, it may suggest a lack of boulders on airless bodies that have diameters of several skin depths. However, many additional factors would need to be accounted for to explore this idea completely.

Very small boulders (≤30 cm) experience only a weak response to thermal forcing, suggesting a threshold below which thermally induced breakdown may occur very slowly (driven primarily by grain-scale effects (Molaro et al. (2015)) or not all. This is consistent with the steep increase in ≤30 cm boulders found at the Chang'E-3 landing site (Di et al., 2016). While the production rate of rocks of varying size, as well as the influence of other weathering processes are also at play, this is encouraging for our results. If we were to consider thermal breakdown only, this would correlate to a minimum strength of ~2.4 MPa to initiate fatigue breakdown. They also report that rocks <10 cm at the Chang'E-3 landing site tend to be clustered together, suggesting in-situ disaggregation. This is consistent with our results showing that slightly larger rocks experience stresses throughout their entire volume, perhaps disaggregating them into multiple smaller pieces. The size-frequency distribution of boulders on Mars (Golombek and Rapp, 1997; Golombek et al., 2003) shows a similar increase in small rocks, though the current presence of atmosphere and presumed wet history of Mars makes it more difficult to compare to our results. Limited statistics are available from Basilevsky et al. (2013) for larger boulders on the Moon (between 2 and 10 m in diameter), which also show an exponential trend. The size-frequency distribution of boulders is also available for Itokawa (Michikami et al., 2010), however as a rubble pile asteroid many of its boulders are relics of formation rather than current day weathering processes. Similarly, boulder statistics on Eros (Thomas et al., 2001) are of limited usefulness, as it is thought that a large impact event created the majority of boulders currently present, and removed or buried older material. The Moon and Mercury are the best environments for understanding how thermal processes effect size-frequency distributions of boulders.

Without more data on the size-frequency distribution of boulders on airless bodies, we turn to damage models to relate stress and environmental conditions to damage accumulation and boulder failure. Part of the challenge to predicting failure is the multi-scale nature of the fatigue process. The engineering literature reveals a wealth of knowledge and models of crack propagation at the microscopic scale. However, these continuum mechanics models are typically of individual crack growth in a specific material or sample are difficult to apply in a broader, more high-level context, where the growth of cracks at multiple scales and change in macroscopic properties of an object (e.g. size, elastic modulus, tensile strength) is the desired outcome. Thus, while crack propagation occurs at a microscopic scale, damage models (such as the Paris Law) rely on macroscopic measurements of breakdown and failure to quantify the thermal fatigue process. These macroscopic damage models are highly empirical and not well constrained for natural processes and materials. Even models that can describe geologic materials are difficult to apply in a planetary context due to the lack of laboratory studies of breakdown in vacuum, and at relevant temperature ranges and thermal cycling periods.

Overall, the work we have presented here highlights the complex nature of the response of objects to diurnal thermal forcing, and emphasizes the need for more research to better constrain thermally induced breakdown rates and regolith production on solar system bodies.

*Acknowledgements: Part of this work was supported by an appointment to the NASA Postdoctoral Program at the Jet Propulsion Laboratory, California Institute of Technology, under contract with NASA, administered by the Universities Space Research Association through a contract with NASA.*

Appendix

A.1 Sensitivity Tests

Several sensitivity tests conducted in order to quantify the impact of model setup and geometry choices are described below. All of these results were found to have a negligible impact on our results.

Width of Regolith Volume: The horizontal size of the model domain relative to the boulder does have an impact on its thermal behavior, as radiation is scattered and exchanged through periodic boundary conditions, which may artificially raise temperatures as if there were nearby boulders. We conducted a test with a 1 m boulder in a model domain of twice the standard width (8$D$). The results showed an increase of ~2

K in the peak boulder temperature relative to the standard case, resulting in a change in peak stress of ~0.1 MPa which is within the uncertainty for stresses.

Mesh Resolution and Uncertainty: For boulders ≤ 10 m in diameter, the finite element meshes were refined such that further refinement changed the value of the peak stress by < 0.3 MPa, which was determined primarily by the effects of the thermal contact boundary layer. The artificial stresses induced in some model runs along this boundary layer, and the differences in peak stress produced by variation in the layer thickness (see below), have a value ~ 0.3 MPa. Limited by this value, we set an uncertainty value for boulders < 15 m at 0.5 MPa, which is slightly higher than the refinement value to account for small changes in peak stress with data output timing interval. In these cases, the mesh in the regolith is coarser than in the boulders, but further refinement falls well below the above thresholds on peak stresses. As boulders become larger, the number of mesh elements needed increases while the desired resolution stays the same. As such, for boulders > 10 m the mesh refinement is primarily is limited by computer memory. In these cases, the mesh was refined as much as possible to allow for computation using available memory resources. In these cases, the uncertainty value is determined by the difference in the calculated value between a mesh element's center and boundary at the time and location of peak stress, which can become significant for meshes that are too coarse. These uncertainties are shown in Figure 10. In these cases, the mesh in the regolith is the same as for the 10 m diameter model.

Thermal Contact Boundary Width: Since the change in temperature across the thermal contact boundary is small, the width of the contact boundary layer ($\delta$) has a negligible effect on the results. A value of $10^{-6}$ m, a reasonable grain size for a basalt boulder, was used for all model runs in this study. A comparison to model runs of a 1 m boulder with a $\delta$ of $10^{-5}$ and $10^{-7}$ shows that all three cases have peak interior stresses within ~0.6 MPa and peak surface stresses within ~0.4 MPa of each other. While not completely insignificant, this effect is not strong enough to impact our results qualitatively. It will have a small quantitative effect, comparable to that some of the material parameters described in section 3.4.

Fixed Regolith Boundary Conditions: We compared the stresses induced in a regolith volume with rigid versus periodic displacement boundary conditions. The symmetry of the problem dictates that the displacement at these edges is zero, regardless of how the boundary condition is defined. Thus, the both average and maximum stresses for each boundary condition showed negligible differences (<0.1 MPa). Thus we have determined that implementing fixed boundary conditions on the sides of the regolith will have negligible effect on our results.

A.2 Solar Disc

The default settings in COMSOL model the incident radiation as a point source at infinite distance, with a specified flux. Accounting for the size of the solar disk in how the flux is defined in our model is important to ensure the calculation of accurate temperatures during sunrises and sunsets. These times of day are often when the highest stresses are induced, and while the absolute change in temperature may be small when the size of the solar disk is accounted for, the change in calculated stresses is non-negligible.

In order to accomplish this, the sun's angular diameter was calculated from its position vector. The local solar elevation was calculated for each point on the geometry, and we adjusted the incidence vector to intersect the center of whatever portion of the disk was above the local horizon for each time step. We then scaled the incident flux by the faction of the solar disk showing above the local horizon. Figure A1 shows the peak stress in a 1 m lunar boulder throughout the course of the day. Stresses induced during both sunrise and sunset are reflected in the profiles because each point is the maximum stress at a given time, taken from any location from within the boulder. The solid line represents peak stress in a boulder where the size of the solar disk is accounted for, and the dotted line for a boulder with the sun as a point source. The latter case shows artificially increased stresses at the most relevant times of day, an effect which would be more pronounced in bodies closer to the sun or rotating more rapidly. We emphasize the importance of accounting for this effect in future studies, particularly for asteroidal surfaces.

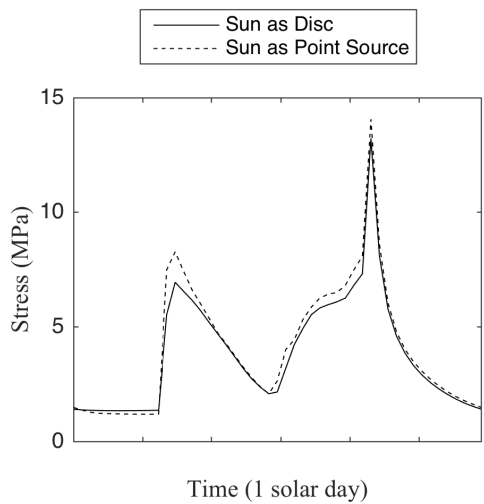

Figure A1. Profile of the maximum principal stress in a 1 m boulder for model runs conducted with the sun modeled as a point source (dotted) and as a disc of finite size (solid). Each point in the profile is the peak stress from any location within the boulder at a given time.